\documentclass[fleqn,10pt]{wlscirep}
\usepackage[utf8]{inputenc}
\usepackage[T1]{fontenc}
\usepackage{bm}
\usepackage{xspace}
\usepackage{textcomp}
\usepackage{svg}

\newcommand{\murm}{%
  \ifmmode
    \mathchoice
        {\hbox{\normalsize\textmu}}
        {\hbox{\normalsize\textmu}}
        {\hbox{\scriptsize\textmu}}
        {\hbox{\tiny\textmu}}%
  \else
    \textmu
  \fi
}

\title{Materials Challenges for Trapped-Ion Quantum Computers}

\author[1,*]{Kenneth R. Brown}
\author[2]{John Chiaverini}
\author[2]{Jeremy M. Sage}
\author[3]{Hartmut H\"affner}
\affil[1]{Duke University, Departments of Electrical and Computer Engineering, Physics, and Chemistry, Durham, North Carolina, USA}
\affil[2]{Lincoln Laboratory, Massachusetts Institute of Technology, Lexington, Massachusetts 02421, USA}
\affil[3]{Department of Physics, University of California, Berkeley, CA 94720, USA}

\affil[*]{e-mail: ken.brown@duke.edu}

\begin{abstract}

Trapped-ion quantum information processors store information in atomic ions maintained in position in free space via electric fields. Quantum logic is enacted via manipulation of the ions' internal and shared motional quantum states using optical and microwave signals.  While trapped ions show great promise for quantum-enhanced computation, sensing, and communication, materials research is needed to design traps that allow for improved performance by means of integration of system components, including optics and electronics for ion-qubit control, while minimizing the near-ubiquitous electric-field noise produced by trap-electrode surfaces.  In this review, we consider the materials requirements for such integrated systems, with a focus on problems that hinder current progress toward practical quantum computation.  We give suggestions for how materials scientists and trapped-ion technologists can work together to develop materials-based integration and noise-mitigation strategies to enable the next generation of trapped-ion quantum computers.

\end{abstract}
\begin{document}

\flushbottom
\maketitle

\thispagestyle{empty}

\section{Introduction}

Trapped-ion quantum computation is based on qubits nearly perfect in isolation, but nonetheless hampered by limited controls and imperfect environments. As ion trap quantum devices increase in complexity, with large efforts at universities, companies, and national laboratories, this is a timely opportunity to consider how improvements in materials can lead to corresponding improvements in the control of atomic qubits and the electromagnetic and vacuum environment of the system. Our review starts with an overview of quantum computation with trapped atomic ions, highlighting recent progress in high-fidelity quantum operations. We then describe the challenge of building integrated ion traps for ion control and measurement. We discuss the materials requirements for combining optical and electrical control compatible with quantum information processing (QIP). We then examine known materials challenge for ion traps, surface electric field noise. We survey the experimental data, describe proposed mechanisms, and suggest future directions for improved analysis and materials. Finally, we discuss ways that ion trap technologists and materials scientists can work together to improve ion trap quantum computers.

There are a number of materials challenges that we do not address here,  including challenges related to generating and manipulating the laser-light necessary for controlling the ions or potential gains of using new materials to improve the vacuum environment, and hence ion lifetimes.

\section{Trapped-Ion Quantum Computation}
\label{IonQC}

\begin{figure}[h]
    \centering
    \includegraphics[width=1.0\textwidth]{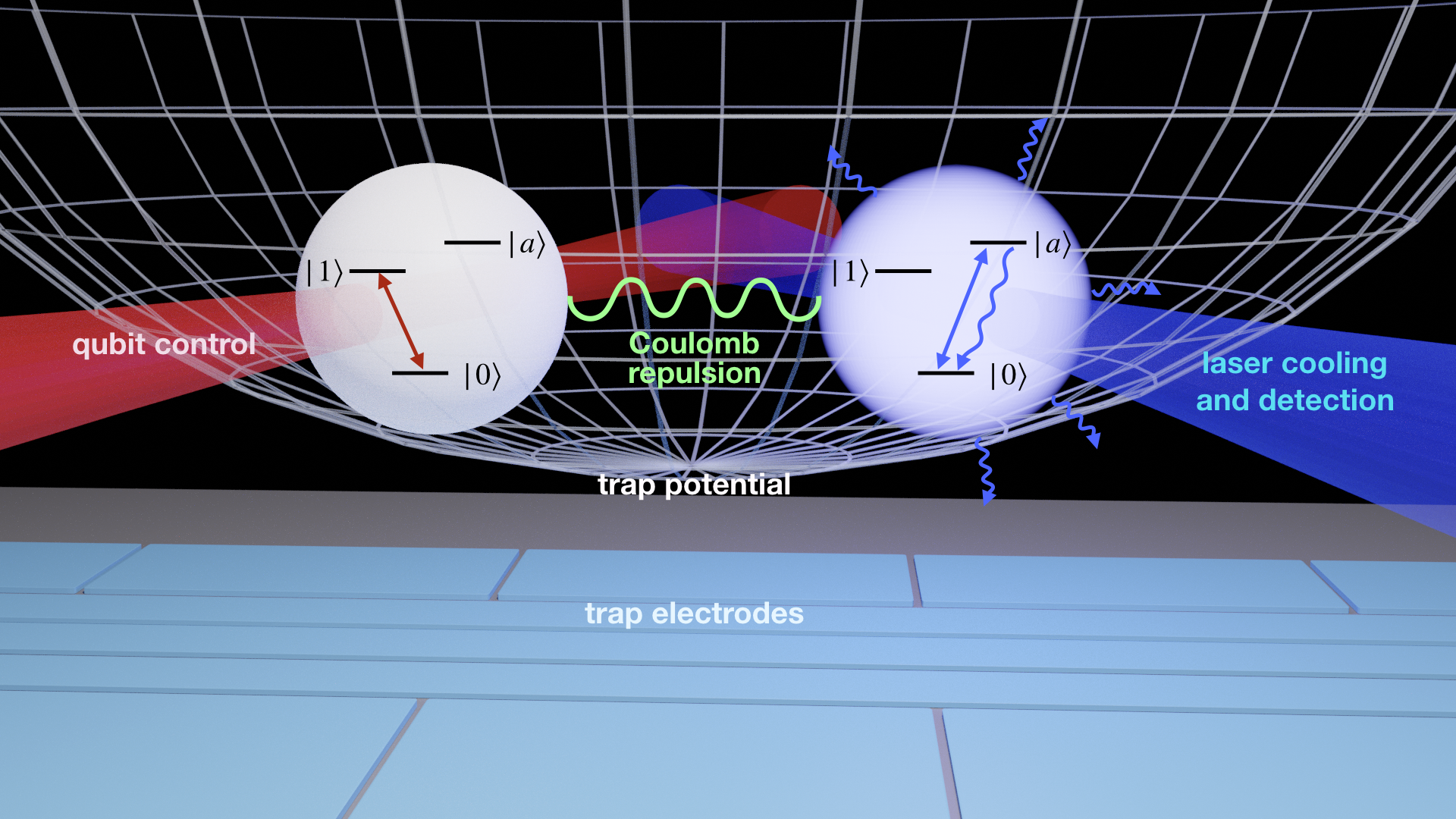}

    \caption{At first glance, there are no materials in an ion trap quantum computer as shown in this schematic. The ions are held in an electromagnetic trap. Lasers or microwaves are used to control the internal states of qubits, $|0\rangle$ and $|1\rangle$. The internal control plus the Coulomb repulsion between ions combine to form conditional logic gates. Readout is performed by measuring laser-induced ion fluorescence using an auxillary state $|a\rangle$. The laser-induced fluorescence  is also used to cool the ions in preparation for quantum logic. The materials challenges arise when we consider how to generate the electromagnetic trap, deliver the control laser beams, and detect the ions' fluorescence. }
    \label{fig:ionbasics}

\end{figure}

A schematic overview of quantum computing with trapped ions is shown in Fig. \ref{fig:ionbasics}. The qubits in a trapped-ion quantum computer are the internal electronic states of individual atomic ions. These internal qubit states are selected to have long coherence times, with reported values of 0.2~s for optical qubits \cite{Bermudez2017} where the information is stored in a superposition of a ground and metastable electronic state, and $>$100~s for hyperfine qubits stored in the magnetic sublevels of the electronic ground state \cite{Bollinger1991,Harty2014,Wang2020arXiv} with dynamic decoupling extending it to $>$6000~s\cite{Wang2020arXiv}.  The ions are held in vacuum by an ion trap. The ion trapping relies on the coupling of radio-frequency (RF) and DC electric fields to the ions' charge, generating an approximately harmonic potential.  This results in ion motion that can be expressed in terms of normal modes, with ``secular'' oscillation frequencies that are typically a few megahertz. 

Control of trapped-ion qubits is achieved through coupling of the quantum states of the ions to classical fields such as  optical and magnetic fields. Single qubit gates are typically driven by optical fields, either via a direct quadrupole transition for optical qubits or a Raman transition for hyperfine qubits.  Hyperfine qubits can also be directly manipulated by microwave-frequency magnetic fields. Single qubit gates of high precision have been demonstrated with fidelities of 99.9999\% for microwave gates \cite{Harty2014} and $\geq$99.995\% for laser-driven gates\cite{Gaebler2016,Bermudez2017}, where the gate fidelity is the probability that, averaged over all input states, the final state is the expected output state.  Two-qubit gates are performed via a combination of applied fields and the Coulombic interaction between the ions.  Optical fields enable the momentum of the photons to transfer motion to the ions in a qubit-state-selective way \cite{Cirac1995a,Sorensen2000}.  For microwave driving fields, an additional static or oscillatory magnetic field gradient is required~\cite{Mintert2001a,Weidt2016,Ospelkaus2011,Sutherland2020}. For optically driven gates, fidelities near 99.9~\% have been demonstrated \cite{Ballance2016-high-fidelity,Gaebler2016} while for microwave driven gates, Bell state fidelities of 99.7~\% have been reported \cite{Harty2016}. 

The fidelity of single and two-qubit gates relies both on the quality of the control of the applied electromagnetic fields and the precision of control of the ion motion. One and two-qubit laser-driven gates suffer from noise in the ion motion, as it reduces the certainty of the ion position, which results in an uncertainty of the laser intensity and optical phase. Two-qubit gates in addition require that the ion motion is coherent for the duration of the gate and ion motional heating on the time scale of two-qubit gates directly limits the ability to entangle ions \cite{Ballance2016-high-fidelity} with high fidelity.  Such heating, which typically arises due to electric field noise, is problematic because it tends to incoherently increase the motional amplitude of the ions, also naturally introducing fluctuations of motion.  Such variation in the excitation of the vibrational modes of the ions not only decohere quantum information present during multi-qubit gates, but also lead to uncertainties in the Doppler effect and thus to ill-defined coupling strengths of laser light to the ions\cite{Wineland1998-experimental-issues}.

Qubit state preparation via optical pumping and cooling of the ions' motion is accomplished primarily by lasers. The measurement of ion qubits is accomplished via laser fluorescence detection; judicious choice of the measurement-laser wavelength is made based on the particular internal level structure of the ion species, such that photons are scattered by the ion only if it is in one qubit state and not the other.  The state of the ion is determined by counting individual scattered photons with a photodector and making a binary decision based on the number of counts in a particular time interval.  The preparation and measurement fidelities correspond to the probability of preparation yielding the correct state and the probability of measurement detecting the correct state. Preparation and measurement fidelities  $\geq$ 99.9\% \cite{Keselman2011, Harty2014, Crain2019, Christensen2019} have been achieved and $>99$\% can easily be obtained. Efforts aimed at increasing the fidelity of all trapped-ion quantum operations, from state preparation and measurement to gates, are ongoing in the research community, and improved performance is expected.

Laser, magnetic, and electric-field  control is typically dynamic; that is, quantum-logic gates and state preparation require amplitude, frequency, and phase modulation of the control fields and ion motion control often requires rapid change of the DC voltages that generate trapping fields, especially quantum computing architectures where high-speed ion shuttling of ions, and splitting and joining of ion chains, are key primitive operations. Furthermore, electric fields oscillating at or near the trap secular frequencies are often used to drive ion motion directly for calibration of trap parameters~\cite{Champenois2001_tickle} and even for the generation of quantum states of motion that may be used as a resource for quantum computing~\cite{Lau2012_IonCVQC, Natarajan1995_motionalsqueezing, Burd2019_motionalsqueezingamp}.  

The currently employed methods and technology for delivery of classical control fields to the ions and for qubit-state measurement are largely based on the availability of the required hardware.  For instance, laser light is typically generated from commercially available laser sources and is delivered to ions from outside of the ion vacuum system via free space and fiber optics.  These optics include mirrors and lenses for passive routing of light, as well as acousto- and electro-optic modulators for active amplitude, frequency, and phase control.  Magnetic fields for quantum-logic gates, oscillating with RF or microwave frequencies, are typically applied via a horn or antenna from the far field, and a DC magnetic field is also typically applied via current-carrying coils in order to establish the quantization axis for the ion's internal spin states. Voltages for setting and varying ion trap potentials and for electric field driving are applied to ion trap electrodes from digital-to-analog converters (DACs) so that they can be computer-controlled at high speed.  These DACs are usually housed in electronics racks, which are remote from the ion trap, and the electrical signals are delivered to the trap electrodes through wires that pass into the vacuum system.   For ion qubit state measurement, scattered photons are typically collected from the ion via a large, free-space, high-numerical-aperture lens~\cite{Crain2019} and these photons are then detected outside of the vacuum system via single-photon detectors such as photomultiplier tubes (PMTs)~\cite{Myerson2008}, avalanche photodiodes (APDs)~\cite{Bock2018_IonAPDdetection, Stephenson2020_REG}, electron-multiplying CCDs (EMCCDs)~\cite{Burrellthesis}, or superconducting nanowire single-photon detectors (SNSPDs) \cite{Crain2019, Todaro2020}.

Universal ion trap quantum computers have been used to demonstrate a variety of algorithms with 2--11 ions \cite{Hanneke2009a, Schindler2013, Fallek2016, Monz2016, Debnath2016a, Linke2017, Hempel2018, Wright2019a,Pino2020} and ion trap quantum simulators with limited functionality have been used to emulate the evolution of spin and spin-boson Hamiltonians using 2--300 ions \cite{Bohnet2016, Zhang2017, Gorman2018}.  These methods all prove the existence of the fundamental building blocks of trapped-ion QIP and demonstrate the incredible improvement and maturation of the field over the past two decades. An important open challenge is how to scale ion trap quantum systems to 1000's of qubits as practical quantum computing is likely to require~\cite{PhysRevA.79.062314}.

Two broad paths towards scalability have emerged among trapped-ion researchers: (1) multiple-site traps based on a ``quantum charge coupled device" (QCCD) architecture with small numbers of ions per site, where interaction is brought about through site-to-site ion shuttling\cite{Wineland1998-experimental-issues,Kielpinski2002CCD,Chiaverini2005,Seidelin2006,Lekitsch2017}; and (2) simpler traps containing larger numbers of ions in linear chains, with multiple such modules joined through photonic interconnects\cite{Monroe2014, Nickerson2014}. There is a great deal of infrastructure common to these approaches, and most proposed systems with photonic links envision a local trap with the ability to order and shuttle ions in a way similar to that required in the QCCD architecture. This naturally requires multi-zone ion traps with small feature sizes in either three-dimensional structures~\cite{Raizen1992a} or on a two-dimensional surface~\cite{Chiaverini2005}. Developing high-quality traps amenable to these architectures while enabling the integration of optical- and electronic-control components and maintaining the capability to perform high-fidelity quantum operations leads to a host of materials-science problems, the main subject of this review.

For further background on trapped ion quantum computation, we refer the reader to high-level views of the technology \cite{Blatt2008}, tutorials\cite{Haeffner2008review, Ozeri2011}, reviews of the current state of the art\cite{Bruzewicz2019,Romaszko2020}, and visions for potential architectures\cite{Kielpinski2002CCD,Monroe2013,Lekitsch2017}.

\section{Materials for Integration of Ion Control and Measurement}
\label{sec:integration}

Conventional ion control and measurement methods and technology has led to impressive fidelity of control and measurement of small trapped-ion systems, but it presents significant challenges to the manipulation of larger systems at comparable fidelity, or even at the improved fidelity that most believe will be required to advance quantum computation beyond the noisy intermediate-scale quantum (NISQ) regime~\cite{Preskill2018}.  One very promising path forward is via integration of classical control and measurement technology into the ion traps themselves~\cite{Lekitsch2017, Bruzewicz2019}.  As chip-scale ion traps are fast becoming a standard trap technology in the field, it is an opportune time to consider on-chip integration technologies such as photonic integrated circuits (PICs) for delivery of light~\cite{Mehta2016}, complementary metal-oxide-semiconductor (CMOS) integrated circuits (ICs)~\cite{Stuart2019}for electric and magnetic-field control and electronic signal processsing, and on-chip single-photon detectors for qubit state readout~\cite{Todaro2020}.

Such integration presents some clear benefits.  For the delivery of laser light to an array of trapped ions, it would be ideal to launch the light from standard optical fiber into on-chip optical waveguides, essentially fiber optics on a chip that define the PIC.  In such a PIC, light could be split up many ways and routed on chip to each location below where an ion resides. An integrated optical element could be placed at the end of each waveguide that converts the light to a free-space mode to address the ions above.  An active optical modulator could also be placed in any waveguide to  allow for independent amplitude, frequency, or phase modulation of ion-control beams.  Not only does this offer great potential for significantly more straightforward and scalable routing of light to an array of ions, but for significantly reduced control noise and drift.  Even in small systems, optical beam-pointing instability and laser phase fluctuations arising from vibrations or drift in bulk optics and opto-mechanics are often the leading source of quantum gate infidelity.  Integrated optics that are physically registered to the ion position, that have much shorter differential beam paths, and that are located entirely in vacuum away from air currents have the potential to reduce this noise considerably~\cite{Niffenegger2020, Mehta2020}.

The collection and detection of photons emitted by ions are essential to trapped-ion QIP, not only for qubit state detection but also for entanglement generation between ions that are spatially separated~\cite{Monroe2014, Nickerson2014}.  For the latter, this may enable trapped-ion quantum networking as well as the modular approach to quantum computing mentioned above, whereby physically separated quantum processing units are connected via interference of photons emitted by ions in each unit.  In all cases, it is desirable to collect and detect photons from ions with as high efficiency and at as high speed as possible.  Additionally, photon-collection and -detection cross-talk needs to be minimized to realize high-fidelity measurement and remote entanglement generation.

The integration of single photon detectors and/or high-NA light collection optics may enable more straightforward measurement and entanglement of large arrays of trapped ions.  For instance, one can envision a detector or light collector located below each ion, as opposed to the conventional approach of using a single high-NA imaging system to collect light from an entire ion array.  By having photon collectors close to the ions, they can be both small and high NA.  For remote entanglement generation, photons collected on chip can be then interfered with one another via PICs instead of relying exclusively on free-space or fiber optics as is done currently~\cite{Hucul2015,Stephenson2020}. Photon detectors placed directly below each ion may obviate the need for collection optics for qubit state measurement altogether, and matching the spatial pitch of detectors to that of the ion array could allow for an efficient use of detector pixels and correspondingly, high-speed simultaneous readout of a large number of ions.

The integration of electronics, and ICs in particular, may also provide significant advantage for trapped-ion QIP. While DACs have been decapsulated and inserted into the vacuum chamber~\cite{Guise2014}, on-chip DACs~\cite{Stuart2019_DACTrap} could dramatically reduce the number of bulk wires that must be attached to apply ion-trap control voltages to the trap electrodes.  This can potentially lead to faster electronic control of ion motion, and it could eliminate a potential source of voltage noise from electrical pickup.  Passive on-chip wiring in general allows for more flexibility and complexity of trap-electrode configuration.  In addition, on-chip wiring could enable local control of the magnetic field across an ion array.  This would be useful, for instance, for locally defining the quantization axis and for shimming of drifting magnetic fields.  A further important application is the generation of large magnetic-field gradients for multi-ion-qubit magnetic-field gates~\cite{Ospelkaus2008,Ospelkaus2011,Lekitsch2017,PhysRevLett.122.163201}, which have the benefit of eliminating spontaneous emission, a fundamental error source present when using optical transitions for these operations.  For such magnetic-field-gradient quantum-logic operations, integration of large-current carrying wires is essential since the gate speed is determined by the size of the gradient that can be produced at the ion location, and thus the closer the wire can be placed to the ion and the more current it can carry the faster the gate can be.  Integration of ICs other than DACs may also prove highly beneficial.  For instance, they can be used to provide signals for co-integrated optical modulators and process signals from co-integrated photon detectors. Combining these integrated technologies could lead to full on-chip processing and enable faster, real-time feedback from qubit detection to subsequent qubit operations, as will likely be needed for quantum error correction~\cite{Steane2006,Chiaverini2004,Schindler2011a}.

\subsection{Materials Requirements for Integrated Technology}
\label{matreqs}
\begin{figure}[t!]
    \centering
    \includegraphics[width=1.0\textwidth]{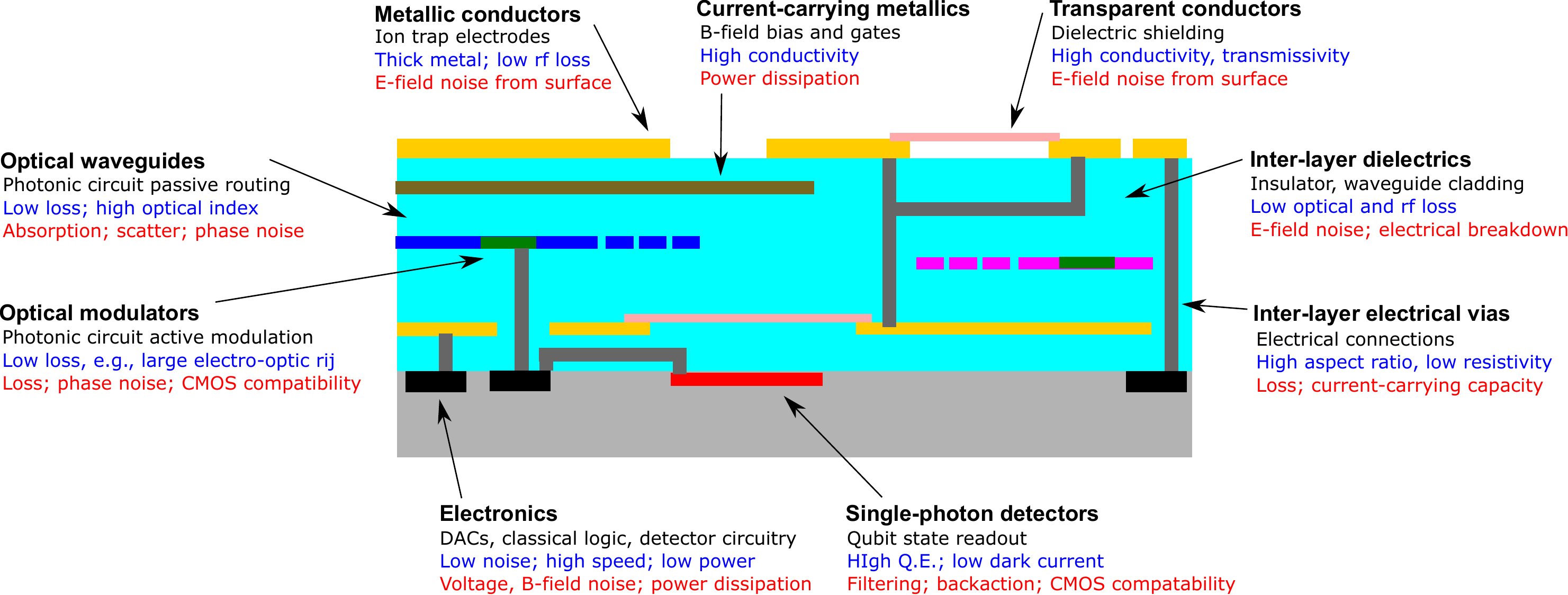}
    \caption{Integrated materials for control of trapped-ion QIP systems.  A schematic cross-section is shown of an ion-trap chip with integrated ion-control and measurement devices and the materials that may be needed for their realization.  Bold black text indicates the device that is needed, normal black text indicates the device function, blue text indicates key required properties of the device material, and red text indicates important concerns that must be addressed for the material to be suitable for trapped-ion QIP systems.}
    \label{fig:mats4integration}
\end{figure}

Integration of ion-control and measurement technology requires that the materials used lead to high functionality of both the integrated devices and the ion traps in which these devices are integrated.  In Fig. \ref{fig:mats4integration} we summarize integration technology that is needed, its required functionality, and the challenges associated with its integration into ion traps.

Materials for PICs for trapped-ion control have a key requirement that they be transmissive over the optical wavelength
range demanded by the atomic structure of the particular ion species in use. For ion species that are typically used in QIP, this wavelength range spans the range
$\sim$300-2000~nm, thus covering most of the near-ultraviolet (UV), to visible, to near-infrared (IR) spectrum.  In addition, to make optical waveguides, at least two different materials are often needed to serve as the waveguide core and cladding, with the core material requiring a higher index of refraction in order to confine and guide the light.  It is particularly challenging to realize low-optical-loss materials for UV and visible wavelengths that are also easily integratable into ion traps.  Indeed, current material choices are dominated by those typically available in CMOS-compatible fabrication facilities, since there are many advantages to leveraging the capabilities enabled by a half-century of Moore's Law scaling.  This enables rapid integration, but also brings along limitations imposed by the relatively small set of materials used in advanced CMOS electronics.  Core materials such as silicon nitride (SiN) and alumina (Al$_2$O$_3$), with refractive indices in the visible spectrum of $\sim$2.0 and 1.7, respectively, and cladding material such as silicon dioxide (SiO$_2$), with an
index of $\sim$1.5, are among those that have proven to have both low optical loss in the UV-visible wavelength range and compatibility with CMOS processing~\cite{SoraceAgaskar2019_SiNAl2O3, West2019_Al2O3}.  
Additionally, to provide tight confinement of the light and make PICs more compact, it is generally desirable to have a core material with a high index contrast relative to the cladding.  High index
contrast also generally leads to more efficient on and off-chip coupling of the light via diffractive grating couplers~\cite{Katzir1977_gratingcouplers}.   Silicon (Si)-core PICs, perhaps the
most mature of the integrated photonics material platforms due to its ubiquity in data and telecommunications applications,
takes advantage of the high index of Si ($\sim$3.4), but is transmissive only at IR  wavelengths~\cite{Rahim2018_SiPhotonReview}; however, it may still be utilized for trapped-ion control in this part of the spectrum. At the other end of the ion-relevant wavelength spectrum, there is a desire to control UV wavelengths from 313~nm down through 235~nm for the light-ion species Be$^{+}$ and Mg$^{+}$\cite{Gaebler2016,Wan2019,Negnevitsky2018,Ball2019,Warring2020,Todaro2020}, and significant work in Yb$^{+}$ qubits depends on high-intensity pulsed laser light at 355~nm~\cite{PhysRevLett.105.090502,Mizrahi2014b,Wright2019a,Nam2019}; it is so far unclear if the above-mentioned material systems will have sufficient transmittivity and power-handling capability at these wavelengths.  Higher-bandgap materials may need to be explored.  Even so, due to a combination of optical loss arising from both material
absorption and Rayleigh scattering due to roughness of the sidewalls of the  waveguides, the optical loss in integrated waveguides is typically much higher than that found in non-integrated bulk or fiber optics.  For Si in particular, two-photon absorption is a known source of loss that limits the amount of optical power that can be carried by a Si-core waveguide~\cite{Rahim2018_SiPhotonReview}. As a result, there is still a great need for further development of improved materials, and potentially improved waveguide surface quality, for routing of light to trapped ions on chip.

In addition to passive routing of light offered by waveguides, active on-chip modulators of amplitude, frequency, and phase are desirable PIC components.  A requirement of the materials used for these modulators, beyond optical transmissivity, is that there be a mechanism to controllably change the material's index of refraction, typically via an electrical signal. To this end, materials with large electro-optic (EO) or piezoelectric (PE) coefficients are desirable, so that the devices can be made fast and small and be driven with low voltages and electrical power. Lithium niobate (LiNbO$_3$),  gallium nitride (GaN), and aluminum nitride (AlN) are three such materials that can have high EO and PE coefficients, are transmissive over wavelength regions of interest for trapped ions, and can be processed using CMOS tools.  As a result, there has been some promising development of integrated modulators using these materials~\cite{Feng2018_GaNmodulators, Xiong2012_AlNmodulators, Zhu2016_AlNmodulators, Wang2018_LiNbO3modulators}, though none have yet to be incorporated into ion traps. Si is another material for which integrated optical modulators have been made \cite{Reed2010_Simodulators}, though these rely on current-induced depletion of carriers from the Si semiconductor to create a change in index, as opposed to a Kerr effect. One of the primary challenges, particularly for modulation of UV and visible light where Si cannot be used, has been the specialized knowledge required to produce and pattern high-quality films of high EO and PE-coefficient materials, and its lack of overlap with the expertise found in the trapped-ion community.  Additionally, the materials under current investigation for modulator development still have undesirably small EO and PE coefficients for realization of few-to-hundred-micron-scale devices, comparable in size to ion traps and sites in trap arrays, that are desirable in order that this component does not limit the size of array cells.  Either novel device designs or new materials with higher coefficients may need to be developed, or an entirely new approach to UV or visible-light modulators (plasmonics, for instance~\cite{Haffner2018_Plasmonicmodulator}) may be required for optically active PICs to be practical for ion traps.

Integrated devices for detection and counting of single photons emitted by ions will require materials that absorb photons at the necessary wavelengths with high efficiency and have low noise properties such that detection events are only registered when a photon is absorbed (i.e, the detector should have low dark count rates).  For ions in wide use for QIP, the wavelength range needed for detectors is $\sim300-500$ nm, but typically only single-wavelength detection is required for a given ion species and the detectors need not be photon-energy resolving.    The primary class of technology for realizing such detectors today is one where a normal metal or semiconductor absorbs a photon, which in turn generates a cascade of electrons in the material that can be measured electronically.  This technology is the basis for photo-multiplier tubes (PMTs), electron-multiplying CCDs (EMCCDs), and avalanche photodiodes (APDs).  For semiconductors, the material bandgap must be small enough to absorb UV-visible photons, and here Si is typically employed.  These devices work well at room temperature, but dark count rates are often observed to drop significantly as they are cooled, improving performance.

A related but different detector technology is the superconducting-nanowire single-photon detector (SNSPD), which produces an electrical signal when a photon is absorbed in the device, momentarily driving normal the superconducting material of which it is made~\cite{SEMENOV2001_SNSPDdevice}.  SNSPDs, made of materials such as niobium nitride (NbN)~\cite{SEMENOV2001_SNSPDdevice}, tungsten silicide (WSi)~\cite{Marsili2013_WSiSNSPD}, and molybdenum silicide (MoSi)~\cite{Slichter2017,Todaro2020} typically have higher detection efficiency and lower dark count rates than PMTs, APDs, and EMCCDs.  However, they must be operated at cryogenic temperatures below 4 K, making their use potentially more challenging if integrated directly in the ion trap.  While Si APDs are perhaps the most ``technologically ready'' device one might consider for large-scale ion trap integration since they can already be made in CMOS~\cite{Rochas2003_CMOSSPAD}, further development of the materials for APDs to increase detection efficiency and lower dark rates would be extremely beneficial.  At the same time, the development of reliable monolithic integration of SNSPD materials such as those mentioned above, or better yet, materials with higher superconducting critical temperature $T_c$, into ion traps could also be a boon to ion-state detection.

Given the maturity of CMOS, Si-based material systems will likely be employed for active integrated electronics such as DACs, detector biasing, pulse-discrimination, and photon-counting modules, and on-chip classical information processors such as filed-programmable gate arrays (FPGAs) or application-specific ICs (ASICs).  That said, other material systems may be of use for integrated electronics.  In particular, ion traps typically require voltage signals, both DC and RF, with amplitudes that exceed the breakdown voltages of standard CMOS devices.  Higher-voltage CMOS processes exist~\cite{Parpia1987_HVCMOS} but ICs in such processes must be made substantially larger to avoid breakdown, making very large scale integration (VLSI) more challenging.  The development and trap integration of alternate materials for integrated electronics, for instance semiconductor materials with significantly larger bandgap as compared with Si (such as GaN~\cite{Mishra2008_GaNElectronics}), may be required for integration of higher-voltage electronics suitable for ion traps.  This may be especially useful for trap RF potential generation, for which voltages in excess of 100~V at ${\sim}10$--100 MHz are typically needed.

In addition to integrated active electronics which rely on transistors, materials capable of carrying high-current ($> 1$~A) DC and AC signals with low power dissipation will be useful for the on-chip generation of magnetic fields.  Thus far, 3--10~$\mu$m thick electroplated gold traces have been used to handle such high currents on ion traps~\cite{Ospelkaus2011, Allcock2013}, but power dissipation has been the primary factor limiting performance.  The development of higher-conductivity materials that can be fabricated into traces integrated into ion-trap chips would potentially allow for faster, higher-fidelity quantum operations on ions via magnetic fields.  Such materials could be lower-resistivity normal metals, or superconductors.  In the case of the latter, superconducting thin-film materials with high critical current density $J_c$ will need to be developed given the currents and relatively narrow-width (${\sim}10$~$\mu$m-scale) ion-trap traces that are necessary.  Superconductors with high $T_c$ would also be beneficial here to obviate the need for very-low temperature operation of the ion trap system.  In addition, low-penetration-depth superconducting materials may be useful for shielding of magnetic fields, via the Meissner effect, from regions of the ion trap where they are not desired.  Such unwanted fields could be generated, for instance, from the high-current carrying wires themselves or from spurious currents arising from time-varying DAC voltages. 

As discussed below in Sec. \ref{sec:SEFN}, materials for the trap electrodes themselves play a very important role in the performance of trapped-ion QC vis-a-vis fidelity of operations conveyed via the Coulomb interaction, though it is still an open question as to what properties these materials should have to optimize performance.  At the very least, trap electrode materials should have high conductivity to minimize power dissipation.  Additionally, as the complexity of trap-electrode geometries are increased to allow for improved functionality~\cite{Maunz2016}, it is almost certain that multiple metal layers will be required for routing, which will further necessitate the use of non-conductive dielectrics to insulate the metallic layers from one another, as well as metallic vias that connect the metallic layers though the dielectrics.  The development of dielectrics with high breakdown voltage will be beneficial to allow for higher voltages on the traps, and hence faster operations.  In addition, the development of dielectrics with low permittivity and low loss tangent at RF frequencies will be useful for improving RF power dissipation in traps, the former being due to the fact that this dissipation scales as the square of the ion trap capacitance~\cite{Siverns2012a}.  

With the potential for integrated photonics being routed underneath the metal trap electrodes, holes in the electrodes will necessarily need to be opened in the metal to allow light to reach the ions. Such holes present a major concern for exposing the ions to dielectrics which can charge up and degrade ion trapping performance.  As a result, the development and integration of materials with high conductivity and optical transparency over the $\sim300-2000$~nm wavelength range may be crucial.  Indium tin oxide (ITO) is one such material that can be used today to cover holes in trap electrodes above integrated photonic components~\cite{Eltony2013_detectorITO, Niffenegger2020}, but it is currently an open question whether this material has sufficiently high conductivity and low electric-field noise for trapped-ion QIP.

In addition to these considerations important for the integration of control technologies in quantum computing systems, related applications have additional requirements.  Atomic sensors and optical clocks are in many cases more useful if portable, and materials used for integration in these systems must therefore have vacuum properties amenable to UHV conditions with reduced pumping capacities, e.g. low out-gassing and bake-ability.  Low power dissipation is also beneficial in such applications, both for lower size and weight, as well as for better temperature control.  In the long run, as quantum sensors utilize more properties traditionally associated with quantum computers~\cite{Kessler2014,Reiter2017}, and as quantum computing nodes are developed for quantum networking applications, the requirements for systems used for all these applications will likely converge, meaning that these additional considerations should be kept in mind in general.

\section{Materials for Ion Traps}
\label{sec:SEFN}

Ion traps for quantum computation are typically made from metal electrodes supported by insulating structures, with recent examples utilizing metal films deposited on insulating substrates, such that advanced microfabrication techniques can be leveraged as described in Sec.~\ref{sec:integration} above.  Such microfabrication also allows for reduction in electrode size, and hence a similar reduction in ion-electrode distance $d$.  Many important performance criteria for ion traps improve significantly when this distance is small. In particular, the strength of the confinement scales with $1/d^{2}$, thereby increasing secular vibrational frequencies for a given voltage, and thus the relevant energy scales and the natural speed of quantum gates. Moreover, small ion-electrode distances speed up splitting and merging operations \cite{Wineland1998-experimental-issues,Kielpinski2002CCD,Home2006-splitting} and ion transport \cite{Walther2012,Bowler2012}, enable large photon collection efficiencies without additional optics \cite{Slichter2017,Todaro2020}, and strong magnetic field gradients for laser-free gates\cite{Mintert2001a,Ospelkaus2008,Ospelkaus2011,Weidt2016}. However, when ions are close to the electrodes, the coherence of the ion motion is compromised by electric field noise generated at materials surfaces.

First measurements revealed significant heating of the ion motion after cooling an ion to the ground state of the trapping harmonic well~\cite{Turchette2000-heating}. The heating rate was orders of magnitude larger than what was expected when considering fundamental limits such as Johnson-Nyquist noise\cite{Johnson1928,Nyquist1928} (JNN).  
The original observations have been confirmed many times over by the community at large; in many instances these large noise values can not be explained by technical noise, such as electrical noise from voltage sources and electromagnetic pickup. Because of the surprisingly large observed noise, the resulting elevated heating rates have been dubbed ``anomalous''. Despite a large body of work, the mechanisms generating the electric field noise are still rather poorly understood \cite{Brownnutt2015}.
Important basic observations, however, are that the electric field noise increases dramatically with the proximity of the ion to the electrode surface \cite{Turchette2000-heating,Deslauriers2006-scaling}, is smaller at large secular frequencies\cite{Brownnutt2015}, and decreases by orders of magnitude when the traps are cooled~\cite{Deslauriers2006-scaling,Leibrandt2009,Chiaverini2014a,Sedlacek2018-multi-mechanisms}.

Early on it was speculated that the surfaces, either those of the metal electrodes or the nearby insulators, play a central role in ion heating via surface-electric-field noise (SEFN). Convincing evidence of this position can be found through treatment of the trap-electrode surfaces.  In particular, a series of experiments  has found that SEFN is reduced when the trap surface has been treated with pulsed-laser excitation~\cite{Allcock2011-heating-reduction}, low-energy plasma~\cite{McConnell2015-plasma-cleaning}, or by large factors when using energetic ions\cite{Hite2012,Daniilidis2014-ion-milling,Sedlacek2018-multi-mechanisms}, i.e. so-called ion beam milling or machining.  There is also evidence that the noise amplitude is not affected by significant changes in the electrical properties of the electrode-metal bulk~\cite{Wang2010}, further implicating the surfaces:  SEFN has been measured to be nominally identical above and below the superconducting transition temperature of the electrodes~\cite{Wang2010,Chiaverini2014a} in cases where JNN was not the limiting noise source.  Thus, while the specific mechanisms of the SEFN behind ion heating are unknown, their main source appears to be the electrode surfaces and possibly adjacent bulk.

\subsection{Experimental observations of surface electric field noise} 
\begin{figure}[tb]

    \centering
    \includegraphics[width=0.7\textwidth]{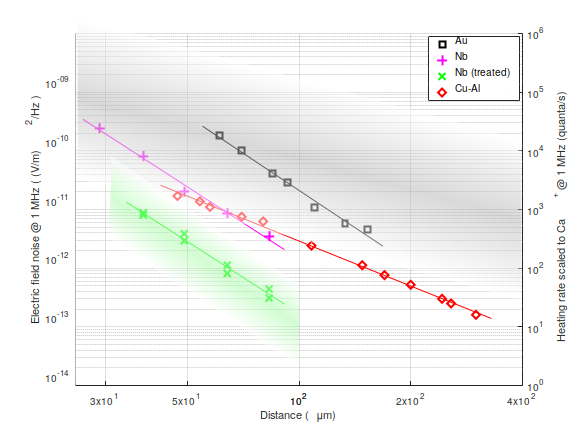}
    \caption{The electric field noise experienced by trapped ions increases with decreasing ion-electrode distance $d$. Only room temperature data from planar traps was selected and  normalized to the expected heating rate of a $^{40}$Ca$^+$ ion at 1~MHz under the assumption that the electric field noise scales as $1/\omega$.
    Black data points were measured above a gold surface \cite{Boldin2018}, magenta above a niobium surface\cite{Sedlacek2018-distance-scaling}, green above a niobium surface treated before installation into the vacuum chamber with ion-milling\cite{Sedlacek2018-distance-scaling}, and red above a copper coated aluminum surface  \cite{An2019-distance-scaling}.  
    Most notably the distance scaling  of the gold and niobium traps is $1/d^4$, while for the Cu-Al trap it is $\sim1/d^{2.5}$. 
    In addition, the black shaded area indicates noise observed in planar gold traps at room temperature, and the green area indicates noise observed in traps treated with ion milling. 
    \label{fig:distance-scaling}}

\end{figure}

In order to learn more about the causes of SEFN, it is helpful to study how the noise spectral density scales as a function of the distance from the material. In particular, we expect that for uncorrelated microscopic noise sources (i.e., when the spatial correlation length is much smaller than the distance $d$ of the ion from the nearest electrode) the noise at the ion position scales as $1/d^{4}$ (Refs. \cite{Turchette2000-heating,Brownnutt2015}). 
Measurements in a single device with a well-defined geometry have only become recently available and seem to verify this expectation \cite{Boldin2018,Sedlacek2018-distance-scaling}. However, a more recent measurement \cite{An2019-distance-scaling} found a power law exponent of $-2.5$, a strong deviation from $-4$. Fig.~\ref{fig:distance-scaling} illustrates these measurements. Shaded areas indicate noise values generally observed in traps with gold electrodes and ion milled traps, though only data from room temperature traps with planar geometry are used in order to facilitate interpretation. It is noteworthy that the measurements in Ref.~\cite{Sedlacek2018-distance-scaling} above a niobium surface and the one in Ref.~\cite{An2019-distance-scaling} above a copper-aluminum alloy show rather small SEFN as compared to other room temperature results, making it likely that both experiments probed a mechanism relevant to state-of-the-art levels of ion heating.

\begin{figure}[tb]
    \centering
    \includegraphics[width=0.57\textwidth]{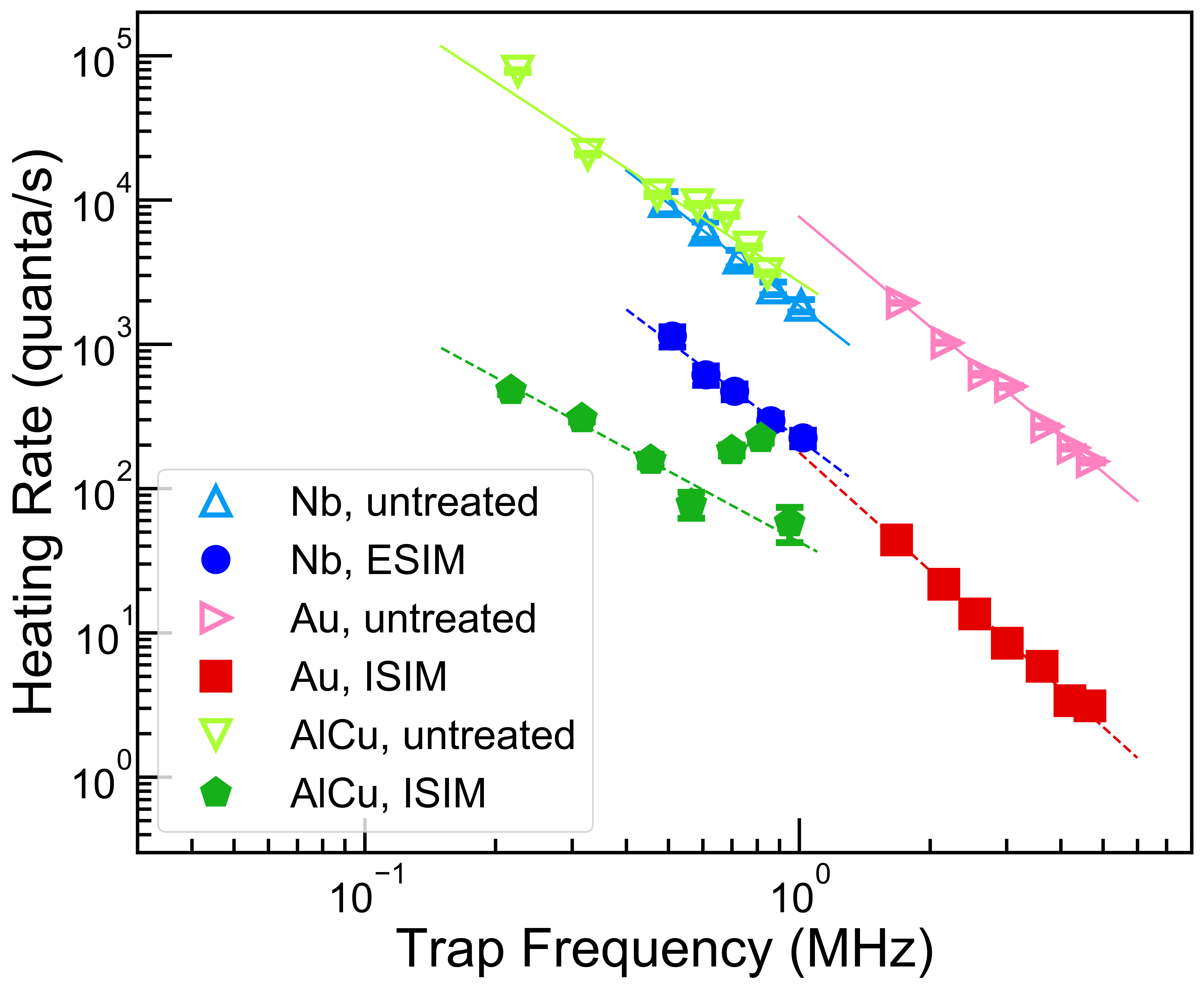}

    \caption{Measured heating rates as a function of trap frequencies before and after treatment via ion milling for surface traps made from three different materials (Nb: Ref.\cite{Sedlacek2018-multi-mechanisms}, Au: Ref.\cite{Hite2012}, AlCu: Ref.\cite{Daniilidis2014-ion-milling}).  All data were taken for room temperature traps.  Heating rates are normalized to a ${}^{40}$Ca$^{+}$ ion at a distance $d=50$~$\mu$m from the trap surface assuming a $d^{-4}$ power law.  As can be seen, while treatment with ion milling reproducibly reduces the heating rate by 1 to 2 orders of magnitude depending on the type of treatment [\textit{in situ} (ISIM) or \textit{ex situ} (ESIM) ion milling], the frequency scaling of the heating rate appears to be unchanged after treatment in each case.  Moreover, all materials show roughly the same frequency scaling; the scaling exponents extracted from the power-law fits shown here vary in the range of $\alpha=0.95(28)$ to $\alpha=1.57(4)$, but the pre- and post-mill exponents are closer to each other for each data-set pair.  While ESIM would seem to bring about a smaller reduction factor, all the treated samples are at approximately the same heating rate level when accounting for the frequency scaling.  The Nb trap, treated with ESIM, was measured in a system that did not require a bake-out, unlike the traps made from the other two materials.
    \label{fig:HRvf_w_mill}}

\end{figure}

In addition to the distance scaling, the scaling of SEFN with trap secular frequency $\omega$ also yields important information about the underlying mechanisms.
Most experiments find that the electric-field noise spectrum scales with some power law, i.e. $S_{E}(\omega) \propto 1/\omega^{\alpha}$ with $\alpha$ ranging between -1 and 7 \cite{Brownnutt2015}. A potential explanation for this huge range is that some of these experiments may have been limited by technical noise sources rather than by the intrinsic noise properties of the ion trap. Indeed, recent, more controlled measurements tend to find $\alpha$ in a much narrower range, i.e. between 1 and 1.5 \cite{Hite2012,Sedlacek2018-multi-mechanisms,Noel2019-TAF}.

While the power-law frequency dependence appears to be close to $1/f$ in nature, the results mentioned above are for untreated traps.  As electrode treatment via ion milling drastically reduces SEFN, one might expect that treated traps would thus be exposing a different mechanism from that seen in untreated traps, and therefore perhaps the frequency scaling would be different.  On the contrary, recent experiments comparing untreated and milled electrode surfaces show rather convincingly that the frequency scaling exponent is both unchanged by milling, and is generally in the same range described above, even when measured on different materials by different groups.  Fig.~\ref{fig:HRvf_w_mill} shows measurements of treated and untreated traps made from Au~\cite{Hite2012}, Cu-Al~\cite{Daniilidis2014-ion-milling}, and Nb~\cite{Sedlacek2018-multi-mechanisms}; the frequency scaling exponent is in the range of approximately $-1$ to $-1.6$, and while there are slight differences in the exponent between materials, each untreated/treated pair leads to equivalent exponents. This comparison also shows the large reduction in SEFN gained from ion-milling treatment, leading to similar levels of noise in treated samples whether the milling is done \textit{in situ}, i.e. in the ion trap chamber,  or \textit{ex situ}, i.e. before inserting the trap into the vacuum chamber.  This similar frequency scaling after treatment of various materials can be used to constrain potential theories of SEFN, as will be addressed below, but it should be mentioned that for the Nb and Au traps the reduced noise levels are still well above what would be expected from JNN while for the Cu-Al trap, noise related to the losses in the dielectric material of the attached capacitors could have contributed significantly to the residual noise\cite{Daniilidis2014-ion-milling}.

Typically, ion beam milling is used to modify metallic surfaces, for instance, by removing contaminants such as carbon and oxygen. Indeed this has also been observed in the above mentioned experiments. This correlation suggests that contamination of the trap surface, such as from carbon and oxygen, is responsible for the noise. While some work supports this causality by linking diffusion of carbon on gold surfaces with density functional theory and scanning probe microscopy to electric field noise \cite{Kim2017Theory-and-Experiments}, other work found that  even with heavy carbon and oxygen contamination, the electric field noise can be rather small on copper-aluminum surfaces (Cu-Al, ISIM data in Fig.~\ref{fig:HRvf_w_mill}) \cite{Daniilidis2014-ion-milling}. Additional work found that removing contaminants via plasma cleaning reduced surface noise; however this did not appear to be as effective as ion milling\cite{McConnell2015-plasma-cleaning}. Thus, it is likely that aspects of ion milling other than removal of contaminants, such as change of the surface morphology, change of chemical bonds, and/or affecting the nature and distribution of defects, play an important role as well.

\begin{figure}[tb]
    \centering
    \includegraphics[width=0.8\textwidth]{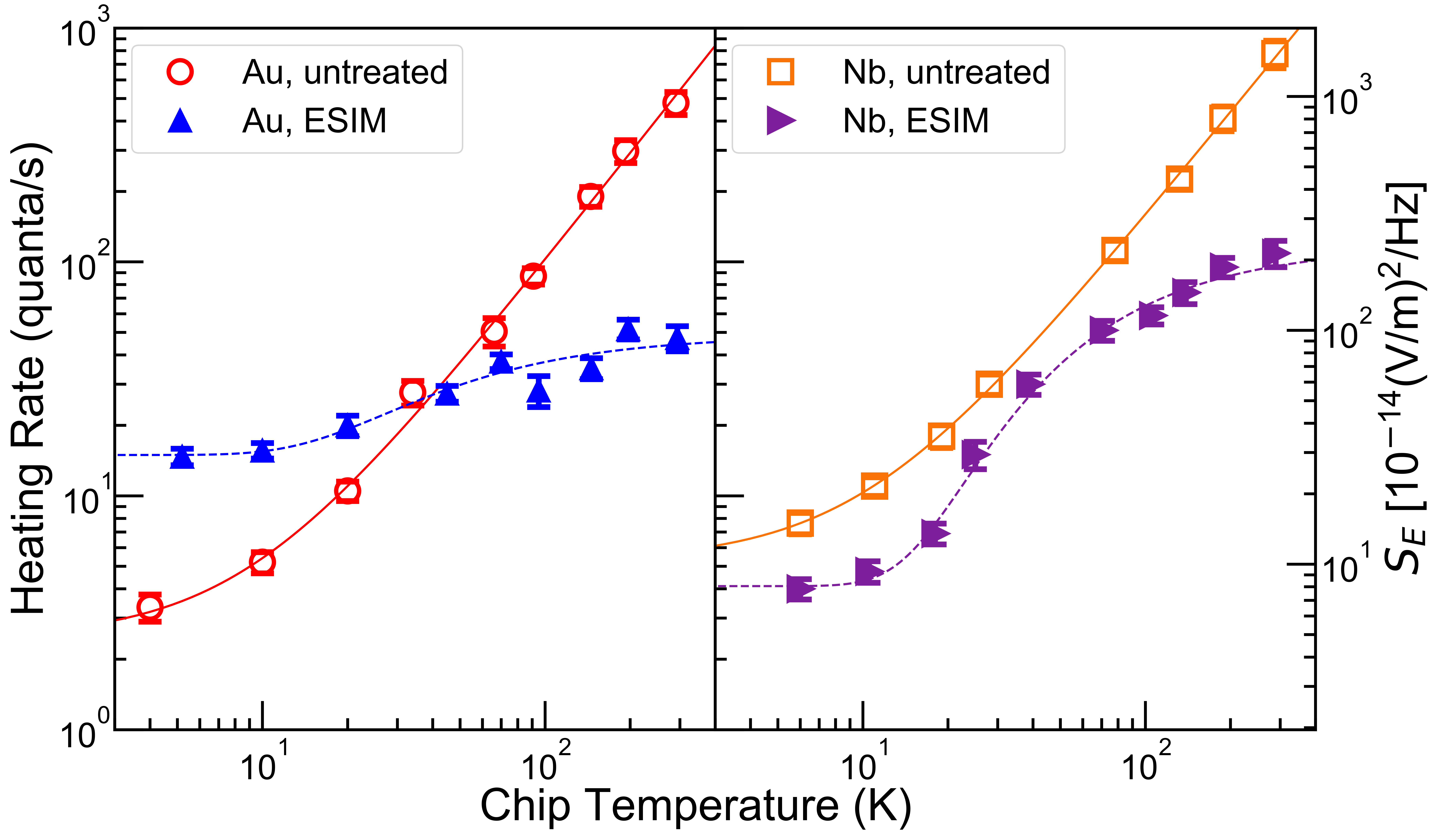}

    \caption{Measured heating rates, as a function of trap temperature, before and after treatment via \textit{ex situ} ion milling (ESIM) for surface traps made from electroplated Au and sputtered Nb (data from~\cite{Sedlacek2018-multi-mechanisms}).  The ion-electrode distance was $50$~$\mu$m and the trap frequency was $1.3$~MHz; $^{88}$Sr$^{+}$ ions were used.  Equivalent electric-field noise spectral density $S_{E}$ at the trap frequency is shown on the scale to the right. Untreated samples show similar behavior, but treated samples display material dependence, even leading to higher electric-field noise levels after treatment of Au at low temperature.
    \label{fig:Au-Nb-mill-temperature}}

\end{figure}

The temperature dependence of SEFN is another key property in ruling in or ruling out particular mechanisms.  Early measurements showed that reducing the trap temperature could reduce the magnitude of SEFN~\cite{Deslauriers2006-scaling}, and subsequent measurements which have extended the temperature range of these measurements have upheld this finding~\cite{Labaziewicz2008}.  However, these measurements showed that untreated traps made from different materials produce very similar SEFN magnitudes and temperature scalings~\cite{Chiaverini2014a}, further implicating surface contaminants.  Interestingly, it appears that removal of some fraction of these contaminants is associated with an uncovering of SEFN material-dependence; the idea of a complicated interaction between specific material properties and effects of ion milling is supported by recent findings that electric field noise above gold is affected very differently than that above niobium subsequent to ion milling\cite{Sedlacek2018-multi-mechanisms}, as shown in Fig.~\ref{fig:Au-Nb-mill-temperature}. In this particular work, it was found that while the electric field noise was very similar throughout the temperature regime from 4 to 300~K for both materials, after ion milling the noise level for gold increased at 4~K while it decreased for niobium.  This suggests much more than a simple additive noise contribution from surface adsorbates, and indeed implicates either other, more subtle, changes effected by surface treatment, or a heretofore unanticipated noise mechanism.

\subsection{Underlying mechanisms and noise mitigation}

\begin{figure}[tb]
    \centering
    \includegraphics[width=0.9\textwidth]{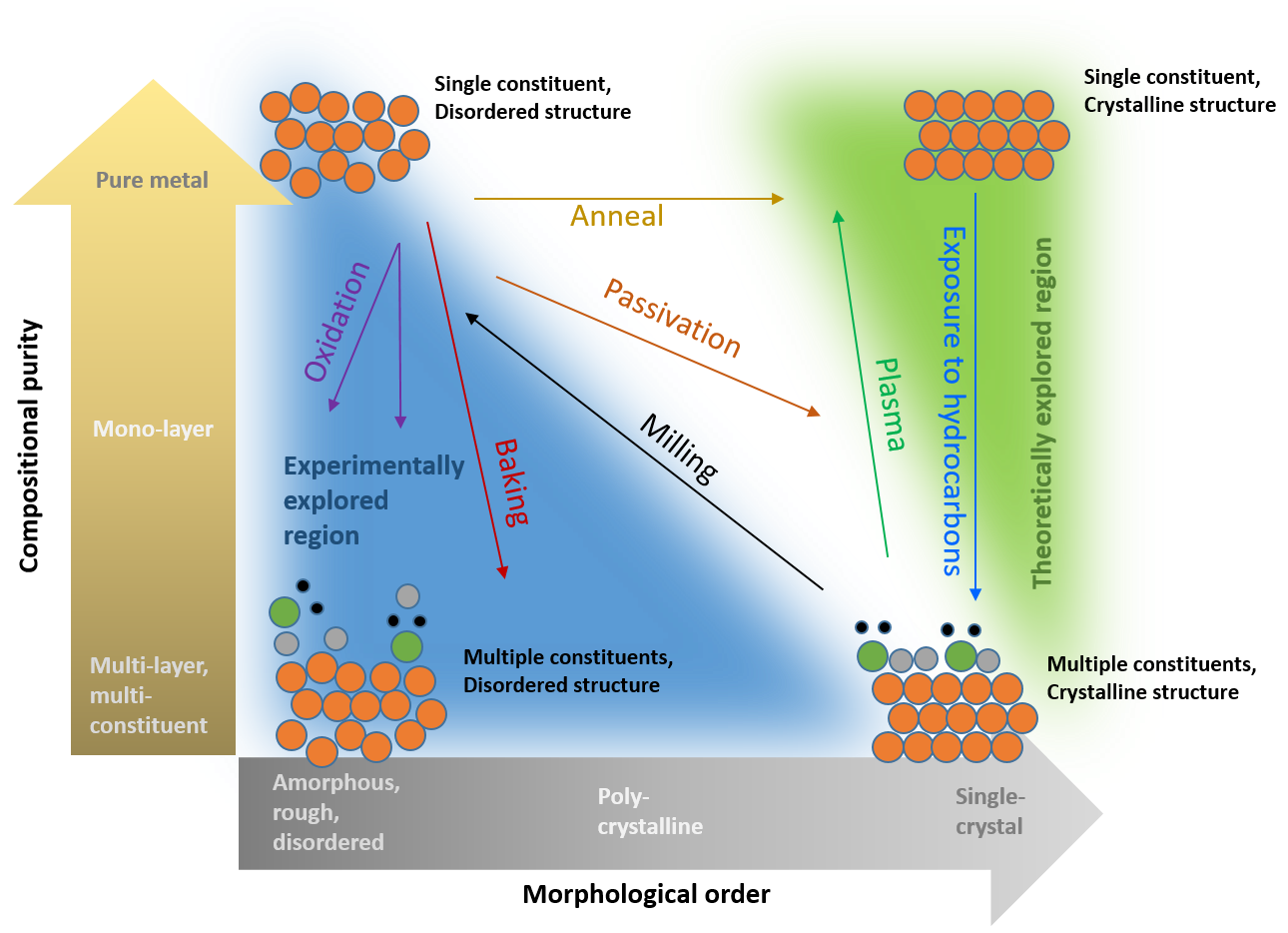}

    \caption{Composition-Morphology space for trap-electrode material surfaces.  Axes represent the amount of order in each of these dimensions, with examples listed at various positions along the axes (large arrows).  The assumed paths taken by a surface in this space under various treatments and procedures is depicted by the narrower arrows.  Cartoon surface cross-sections in four quadrants of the space are meant to be suggestive of potential atomic and molecular structure and composition at these extreme points.  Structure and paths are meant to be notional. 
    \label{fig:comp-morph}}

\end{figure}

There is much we can say about SEFN, but in terms of determining how materials ultimately can be tailored for high-performance trapped-ion quantum computing, much is still unknown.  Seemingly chief in importance, among the results from the field over the last several years, is the relevance of the electrode (and potentially exposed insulator) surfaces, and the large effect of surface treatment on measured heating rate. Despite these encouraging results, it is still not clear what the different treatments do to the surface at a microscopic level.  Moreover, while some effort is being made in this direction~\cite{Hite2017b}, it is still rather unclear what the surfaces of untreated electrodes look like at this level.  And perhaps paramount in efforts to mitigate SEFN, it is not understood what we \textit{want} those surfaces to look like microscopically in order to produce the best results.  Surface science and condensed matter research has typically striven to produce very clean, very ordered surfaces in order to simplify applicable models and reduce the supposedly deleterious effects of disorder.  On the other hand, some amount of disorder can lead to interesting and important coherent effects in materials, in some cases leading to lower noise.  It is thus not straightforward to even know what to aim for.

Fig.~\ref{fig:comp-morph} attempts to describe the resulting path taken in a two-property space by various operations performed on ion-trap surfaces as part of typical experiments to date.  This includes not only the purposeful treatments described above, but also procedures encountered during or after fabrication and exposure to solvents or atmosphere, and procedures that are routinely applied to metals on substrates during surface and materials processing. Composition, i.e. the makeup of the surface, nominally in terms of constituent atomic of chemical species, is clearly important to the behavior of many solid-state systems.  Likewise morphology, i.e. the structure of the surface quantified by roughness, long-range crystalline order, etc., is fundamental to material behavior.  As both of these properties are altered during processing, exposure, and treatment of ion-trap surfaces, this phase-space may provide some guidance for experimental campaigns seeking to expose SEFN mechanisms.  Once again, while states of high compositional and morphological order at the trap surfaces, as in the upper-right of the diagram, may appear desirable, is is not known \textit{a priori} that this will also result in low SEFN.  In fact, measurements of surfaces more accurately categorized as being in the middle to lower part of this diagram, near the center on the Morphological axis, have resulted in SEFN among the lowest observed.  In particular, ion milling is known to increase morphological disorder, and the decrease of room temperature heating rates after ion milling even after re-contamination~\cite{Daniilidis2014-ion-milling,Sedlacek2018-multi-mechanisms} points to the the left of the diagram as a desirable region. The observed SEFN in both cases is still above fundamental levels, however, so it is difficult to say that the upper-right area, minimally explored to date, will not ultimately lead to noise levels near the JNN limit.

A strategic approach may be to attack the problem along one axis at a time while controlling the level of order along the other axis, such that the role of each may be isolated.  This will require effort in understanding more precisely the paths taken in phase space by various procedures, a task likely most efficiently undertaken with input from surface scientists.  Such expertise can be leveraged not only to determine these paths, but also to suggest alternative methods of exploring this space more orthogonally, as well as to suggest alternative measurements that can indicate where in this phase space one is and would want to be. 
Additionally, it is likely that this phase-space picture is incomplete, since there are other variables that may affect SEFN.  For instance, properties of the electrode metal likely become important after removal of all contaminants. Thus the ideal surface configuration for the elimination of SEFN could also depend on properties such as conductivity, reactivity, etc.

\subsubsection{Proposed mechanisms}

There are many properties about SEFN which are known.  These can help to constrain the theories that have been put forth in order to explain SEFN, but they, in combination with available theories, do not yet paint a complete picture of what causes  SEFN and what can be done to (completely) mitigate it.

Several phenomenological mechanisms have been suggested, and they can each reproduce some subset of the observed scaling behaviors described above.  These models include patch-potential models, based on electrode surfaces composed of a collection of patches uniform in size, each with an independent time-varying potential\cite{Sandoghdar1992,Turchette2000-heating,Dubessy2009,Low2011}.  The correlation length of the potentials on these patches compared to the distance $d$ from the ion to the electrode surface, as well as the geometry of the surface, predict distance scalings of electric-field noise, and for patch size small compared with $d$, the typically observed $d^{-4}$ scaling is recovered, as mentioned above.  Unfortunately, this model is merely a framework, and without a mechanism for the potential variation across patches, scalings with other relevant quantities are not predicted.  Another phenomenological model posits that loss in dielectric layers above the electrode metal surfaces leads to noise at the ion location~\cite{Kumph2016}.  Thin adsorbate layers are common on untreated ion-trap electrodes, and this theory predicts realistic levels of electric field noise at room temperature for reasonable estimates of the thickness and dielectric properties of these adsorbate layers~\cite{Kumph2016}.  It also predicts $1/f$ scaling with frequency and $d^{-4}$ distance scaling, but it is difficult to compare precisely without better knowledge of the dielectric properties of hydrocarbon residues as a function of temperature, not to mention the actual thicknesses of these layers and whether or not this dielectric-layer assumption breaks down for few-monolayer levels of contamination, levels that have been seen experimentally~\cite{Kim2017Theory-and-Experiments}.  A third category of phenomenological models is that based on thermally assisted fluctuations (TAF) of two-level systems at or near the electrode surface~\cite{Noel2019-TAF}. While no particular microscopic two-level system is implicated, the general $1/f$-like nature of the scaling resulting from a distribution of fluctuators with varying energy splitting, and the coupling of temperature and frequency scaling that ensues, has been observed in some recent experiments~\cite{Noel2019-TAF}.  The distance scaling is not predicted directly from this model, unless patches or some other geometry is assumed.  These phenomenological models all show promise, but all require additional theoretical or experimental input to make specific predictions.

There have also been more specific models proposed to explain SEFN, where a microscopic mechanism, or combination thereof, is posited.  Overall, however, these models all generally fail to predict some key aspect of the state of reproducible measurements, described above.  It is instructive to see where they fall short, however, so that differences from observation may potentially guide modification based on more realistic assumptions.  As individual contaminant atoms and molecules will exist on any surface, even in ultra-high vacuum conditions, one might expect these contaminants' presence and motion to lead to electric field noise.  In fact, both fluctuations in the dipolar field produced by adatoms bound to a surface~\cite{Safavi-Naini2011,Safavi-Naini2013}, and adatom diffusion across electrode surfaces~\cite{Brownnutt2015,Kim2017Theory-and-Experiments} have been suggested as potential mechanisms.  Both theories predict an activated (Arrhenius) type scaling with temperature---this is similar to what is seen in the Au and Nb after treatment with ion milling, data shown above in Fig.~\ref{fig:Au-Nb-mill-temperature}.  While the adatom dipole model predicts a $-4$ distance scaling exponent, in agreement with most experiments, it predicts a flat frequency response near ion-trap-relevant frequencies in the megahertz range, in contrast to experiment.  A recent extension of this theory, using more accurate modeling of the adsorbate-surface interaction~\cite{Ray2019}, has resulted in the same frequency scaling for the relevant frequency range, but the regime of $1/f$-type behavior is closer in frequency to the experimentally relevant megahertz range.  Conversely, the adadtom diffusion model predicts $1/f^{2}$ frequency dependence, not far from what is seen experimentally.  However, this model predicts a distance scaling of $d^{-6}$, a stronger dependence than seen in any experiment.  A modified diffusion model, in which the adatoms diffuse over patches of varying potential, leading to a varying dipolar field with position, has also been suggested~\cite{Brownnutt2015}, and while the predicted frequency dependence of $1/f^{1.5}$ is right in line with some of the data, this model also predicts $d^{-6}$ distance scaling for modes of ion motion parallel to the electrode surface,  unsupported by the data~\cite{Boldin2018,Sedlacek2018-distance-scaling,An2019-distance-scaling}.  Moreover, all these specific models underestimate the magnitude of the SEFN seen in experiments:  either the expected dipole density or strength is too small, or measured diffusion constants are too small, to reproduce the level of noise seen in the laboratory for modes parallel to the surface.

All the models described above are based on equilibrium processes, where the noise is ultimately driven by thermal excitations.  One may reasonably assume, however, that since the operation of an RF Paul trap requires application of high-voltage at RF frequencies, there is some out-of-equilibrium excitation present in the electrode or insulator surfaces and bulk in most trapped-ion experiments.  If this were to be part of a mechanism for the production of electric-field noise, one would expect significant SEFN dependence on RF amplitude; this is not generally seen in experiments where technical noise is not dominant~\cite{Sedlacek2018_tech_noise}.  It is possible that the RF potential leads to saturated excitation for all practical levels of RF voltage amplitude---experiments meant to test this hypothesis are proposed, but there are no developed models positing this mechanism.

\subsubsection{Potential avenues of exploration and mitigation of surface electric field noise}

As none of the current body of mechanistic theories accurately predicts a comprehensive subset of the measurements of SEFN when considering both absolute magnitude of noise and scalings of this noise with relevant parameters, new models, or modification of existing models, is required.  Potential avenues of inquiry include extending adsorbate modeling from elemental and simple molecular constituents to more complex, mixed-species surface coverings, such that lower-frequency modes of adsorbate motion in the surface potential can be effectively modelled~\cite{Ray2019}.  Beyond pushing frequency dependence into a more experimentally relevant regime, this approach is consistent with the types of adsorbates likely present after trap fabrication, notably solvents and polymeric hydrocarbons like photoresist.  Alternatively (or better, in conjunction), more precise modeling of two- and multi-level systems present at and near surfaces should be undertaken where possible, as the rich behavior of collections of such systems is unlikely to be reproduced with the simpler models constructed to date. More specific guidance can also be obtained if one manages to connect to the vast knowledge about two-level fluctuators in bulk or by specific measurements identifying the nature of two-level systems near surfaces using surface characterization tools. There are of course practical limitations to general modeling of noise, stemming from computational complexity and a lack of knowledge of underlying adsorbate identity and behavior at the molecular level.

Beyond new models, new experiments are also suggested.  Since exact modeling of the presumed conditions of ion trap electrodes and environments is challenging, experiments in more limited scenarios, e.g. with less complex surface characteristics, could be greatly beneficial in narrowing the breadth of possible areas of weakness of particular theories.  By the same token, experiment-model mismatch may occur due to actual conditions experienced by trapped ions being different from those presumed models (cf. Fig.~\ref{fig:comp-morph}).  Thus more targeted, and more varied, surface and material characterization techniques can shed light on what the experiments are probing with more veracity, potentially allowing for more straightforward comparison with theory. 
In this context, set-ups integrating surface science tools into ion trap apparatuses are of high interest\cite{Hite2017b,Daniilidis2014-ion-milling}. Additionally, almost all theories suggested to date depend on material properties of the electrode material and/or adsorbates, so further exploration by means of experiments sensitive to material dependence~\cite{Sedlacek2018-multi-mechanisms} should be pursued, nominally by exploring more materials such that material properties may be correlated with measured noise levels. Thus, impedance-matching experiments with theory that can be precisely performed to close the experiment-modeling gap is just as important as improving theory.  Both are needed as part of a holistic approach to solving, or at the very least mitigating to useful levels, the problem of SEFN.

Synthesizing current knowledge of SEFN with the discussion of materials considerations relevant for integration of control technologies for practical trapped-ion QIP from Sec.~\ref{sec:integration}, we attempt now to summarize the current knowledge on minimizing the effects of SEFN on quantum logic as a practical guide to mitigation strategies.  Gold, a widely used ion-trap electrode material, has high conductivity and is resistant to oxidation.  Measurements have exhibited a wide distribution of SEFN values extending from state-of-the art to orders of magnitude larger values. Interestingly a state-of-the art trap showed an increase in noise at low temperature after ion-milling treatment~\cite{Sedlacek2018-multi-mechanisms}.  A drawback to gold is its incompatibility with standard CMOS-facility fabrication, though if electrode-metal deposition can be the final step in the process, gold can be incorporated in a separate facility.  Aluminum and aluminum-copper alloys are, on the contrary, widely available in CMOS microfabrication processes, but oxidation does occur and could be a source of SEFN in light of the lossy-dielectric model. Still Cu-Al traps show state-of-the art SEFN noise levels, often lower than Au. Niobium has also been studied and shown to have state-of-the art SEFN levels, especially after (\textit{ex situ}) ion milling.  This is despite the oxides that readily form on niobium surfaces~\cite{Grundner1980}.  In addition, Nb becomes superconducting below approximately 9.2~K, but no difference in SEFN has been observed when crossing this transition~\cite{Wang2010}.  However, superconducting electrode materials may have benefits to integration of other control technologies, as mentioned in Se.~\ref{sec:integration}.  Heavily doped silicon can be used as an electrode material~\cite{Britton2008,Britton2009,Li2017}, and indeed it may lend itself to experiments that can guide understanding of SEFN mechanisms, due to its well-characterized surface and structure; to date, this material has not been explored in detail for trapped-ion quantum computing. 
Finally, large
reductions in SEFN have been observed when cooling ion traps from room to cryogenic temperatures.  This, in combination with the additional general benefits to cryogenic operation, including long ion lifetime~\cite{Antohi2009} and lower resistivity of conductors, suggest its wide use for trapped-ion QIP.

.

\section{Outlook and Future Perspectives}

\begin{figure}[tb!]

    \centering
    \includegraphics[width=1.0\textwidth]{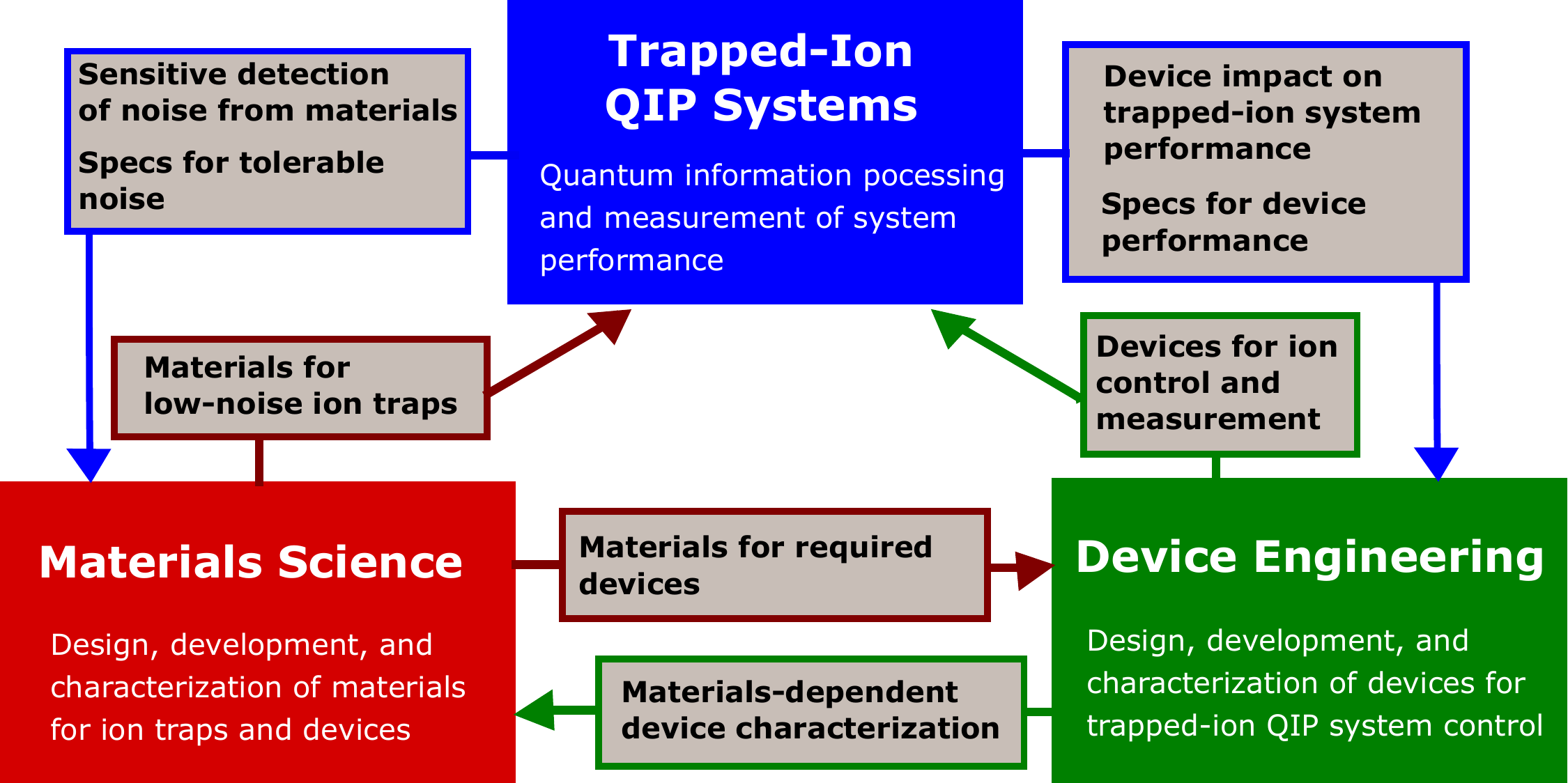}

    \caption{Interconnected fields of study for trapped-ion materials science}
    \label{fig:TIMS}
\end{figure}

With the many desired integrated technologies discussed in Sec.~\ref{sec:integration}, and the desired low-SEFN properties described in Sec.~\ref{sec:SEFN}, there will naturally be performance interdependencies and trade-offs  among the materials required to realize them.  For example, PICs may require high-refractive index materials that come with high permittivity, thus making them less ideal for RF power dissipation.  Or the highest conductivity trap metals may not necessarily provide the lowest electric field noise. Low dissipation insulators are often not easily deposited as part of a CMOS-compatible fabrication process, and exposed dielectrics, potentially required for photonic integration, may be an additional source of electric field noise.  As a result, careful consideration of these material interdependencies, looking from the perspective of the full trapped-ion QIP system, will be required to guide the development of optimal materials.

Nonetheless, the benefits of integration are clear, especially if it can be shown that the integrated devices themselves have performance on par with, or exceeding, that of conventionally used technology. Given the potential it offers, it seems such integration will be a major focus of future work in the field of trapped-ion QIP.  It must be remembered that control-technology integration also presents a significant challenge beyond the development of devices on their own.  One of the great benefits of trapped-ion qubits is their long coherence times, afforded by how well they can be isolated from their environment.  This isolation results primarily from the fact that ions can be confined in a vacuum; however, integration of control and measurement technology necessarily brings materials in closer proximity to the ions, and with this comes the risk of a significant reduction of this isolation.  As a result, it is crucial to study the impact of the materials required for ion trap integration on ion-qubit performance. Materials science for ion trap integration will involve experts in three distinct disciplines: trapped-ion QIP science, device engineering, and materials science.  In isolation, these fields cannot hope to tackle this important problem comprehensively.  In Fig.~\ref{fig:TIMS}, we illustrate the kind of interdisciplinary approach that will likely be required to tackle this problem.  In general, the needs of trapped-ion QIP systems dictate the performance specifications of the devices for control and measurement. These specifications in turn dictate what materials must be developed and used in the devices.  And the properties of the materials and how they impact the nearby ions will ultimately determine their suitability for integration.

This discussion of materials challenges for ion trap integration has focused on the particular application of quantum computation.  However, these integrated systems will have further appeal to building next-generation portable atomic clocks, advanced quantum-limited sensors, and miniature laser-cooled mass spectrometers with potential biological applications. 

To date, study of ion trap materials has been undertaken almost exclusively by atomic physicists, and the narrowness of this field of experts has likely limited the progress of useful system development. We hope that this review article will spark new collaborations between trapped-ion quantum information scientists, device engineers, and materials scientists to address the critical materials challenges that lie ahead for trapped-ion systems.

\noindent\textbf{Acknowledgements}\\
The authors thank Mark Kuzyk for assistance with Figure 1. KRB and HH acknowledge support from the National Science Foundation STAQ project Phy-1818914.  This material is based upon work supported by the Department of Defense under Air Force Contract No. FA8702-15-D-0001. Any opinions, findings, conclusions, or recommendations expressed in this material are those
of the authors and do not necessarily reflect the views of
the Department of Defense. \\

\noindent\textbf{Author contributions}\\
KRB, JC, HH, and JS all contributed to the writing of the manuscript and the preparation of the figures. 
\\

\noindent\textbf{Competing interests}\\
KRB is a scientific advisor for IonQ, Inc. and has a personal financial interest in the company. \\\

\end{document}